\documentclass{article}

\usepackage[utf8]{inputenc}
\usepackage[review]{tismir}
\usepackage{amsmath}
\usepackage{hyperref}
\usepackage{url}
\usepackage{graphicx}
\usepackage{booktabs}
\usepackage{lipsum}
\usepackage{xcolor}

\title{Generating Music with Structure Using Self-Similarity as Attention}
\author{%
Sophia Hager,%
~Kathleen Hablutzel,%
~and Katherine M. Kinnaird}

\date{}

\begin{document}

\twocolumn[{%
\maketitleblock
\begin{abstract}
Despite the innovations in deep learning and generative AI, creating long term structure as well as the layers of repeated structure common in musical works remains an open challenge in music generation. We propose an attention layer that uses a novel approach applying user-supplied self-similarity matrices to previous time steps, and demonstrate it in our Similarity Incentivized Neural Generator (SING) system, a deep learning autonomous music generation system with two layers. The first is a vanilla Long Short Term Memory layer, and the second is the proposed attention layer. During generation, this attention mechanism imposes a suggested structure from a template piece on the generated music. We train SING on the MAESTRO dataset using a novel variable batching method, and compare its performance to the same model without the attention mechanism. The addition of our proposed attention mechanism significantly improves the network's ability to replicate specific structures, and it performs better on an unseen test set than a model without the attention mechanism. 
\end{abstract}
}]

\section{Introduction}\label{sec:introduction}

Recent advances in hardware and in machine learning have made music generation leveraging large datasets through deep learning possible. One of the challenges that remains in data-driven music generation, despite the continued growth of deep learning, is mimicking in a generated piece the repeating patterns present in musical pieces. These could be as simple as verse/chorus structure in pop music, or include repeated themes and motifs as seen in classical music. These patterns and structure in music can be difficult to replicate using basic models that do not intentionally try to mimic these structures \citep{directions}. 

A method to address this limitation is attention \citep{survey}. Traditionally, an attention layer in a neural network provides weights for previous timesteps, allowing the network to be more heavily influenced by some previous steps than others. This often helps provide the repeating structure usually missing from models that do not use attention.  Transformer models \citep{og_transformer}, a commonly used model in modern music generation systems, solely use an attention mechanism for learning sequences. However, like many deep learning layers, attention layers are usually composed of a set of weights that need to be tuned. Additionally, many attention models have very limited explainability. 

\begin{figure}[ht]
    \centering
    \includegraphics[width=0.5\textwidth]{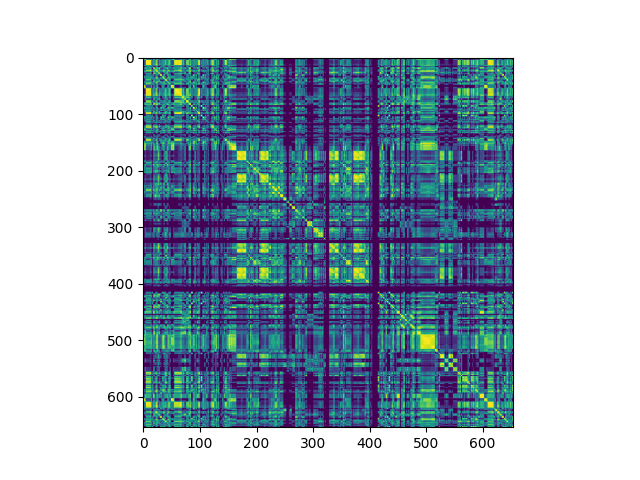}
    \caption{An example of a self-similarity matrix after the preprocessing for batching has been applied. Yellow regions indicate higher similarity, while blue regions indicate lower similarity. There are high level structures, such as the region from about 170-400 having relative similarity to itself compared to surrounding regions, and lower-level structure, such as the more minor variations in similarity within that region.}
    \label{fig:example-ssm}
\end{figure}

One representation of musical structure is the self-similarity matrix (SSM) \citep{ssm}. This commonly used representation in music information retrieval encodes both large and small-scale structure, and can be used as the precursor to other structure representations. An example of an SSM is shown in Fig.~\ref{fig:example-ssm}. In effect, the $i^{th}$ row of the SSM reports which beats the $i^{th}$ beat is closest to (or furthest from). Put another way, the  rows of the SSM implicitly encode attention for each beat.

We propose a method to control the structure of an output piece of music by creating an attention method using the input SSM itself as weights that the network uses for previous timesteps. In this work, we are focusing explicitly on the strengths of using the SSM as the attention layer for music generation, even when--and especially when--we use a less powerful deep learning music generation system like one based on LSTM instead of a transformer. We demonstrate the efficacy of using an SSM for attention by creating a music generation system, \emph{Similarity Incentivized Neural Generator} (SING), that consists of a Long Short Term Memory layer and our SSM-based attention layer. We show that SING generates music that contains similar elements of structure to the input SSM, while a system without the SSM-based attention layer fails to create similar repeated structures. These generated pieces maintain observable structure at longer lengths than other existing methods for structured generation. 

We evaluate SING's performance in two ways: computationally, using a comparison metric, and qualitatively, via a human study. 
Our two evaluations seek to address two different aspects of ``performance.'' 
Our computational evaluation works to understand what gains quantitatively have been made in creating structure in a music generated piece using the generated SSM. 
In our human study, we are investigating what gains have been made--if any--to the sound of the music as experienced by a human listener. 
In both metrics, SING compares favorably to a simple LSTM, meaning that we have a proof of concept for controllable generation with larger-scale models using SSMs as templates for generation (computational evaluation) in addition to better musical quality (human study). 

To train SING, we used a dataset with wide variation in the lengths of the included pieces. Like most deep learning systems, SING requires batching to be tuned and a second contribution of this paper is the variable batching method that we use to train SING. In traditional batching methods, long pieces are split into multiple smaller portions, and short pieces are excluded or extended to fit the standard length.
For example, with a standard length of 400 samples, a 1000-sample piece would be split into two pieces of length 400 each, with a leftover 200 samples to be excluded or extended into another piece. 
These splits, extensions, and exclusions present challenges when training to generate long-term structure. Splitting long pieces into short segments restricts the model from seeing the longer structure of the piece, extensions introduce artificial changes to the piece structure, and exclusions reduce the amount of available data. 

When datasets present large variation in piece length, it can be challenging to select a single piece length that both captures long-term structure and avoids excessive exclusion or extension of shorter pieces in the dataset. Our proposed variable batching method balances the size of each batch\footnote{Limiting the size of each batch was necessary in our case due to hardware limitations. Large batches we split into two, again due to our particular hardware limitations.} while also minimizing the amount of edits (either truncation or padding) applied to each piece.

\section{Motivation and Background}
Computers have been used to algorithmically generate music for over 50 years, beginning with simple melodies generated using Markov chains \citep{survey}. Today, deep learning is widely used to autonomously generate music. Previously, due to the time-based nature of music, much of the deep-learning based generation methods utilized Recurrent Neural Networks (RNNs). In particular, Long Short Term Memory networks (LSTMs) \citep{hochreiter} were often used in music generation systems \citep{survey, grad_prob,lstmsystem}.

Generation systems became more successful over long sequences when attention became commonly used in conjunction with RNNs. Attention creates varied influence on the current timestep using weights, rather than each previous timestep having equal influence \citep{survey}, promoting the development of long-term structure. For example, Google Magenta \citep{magenta} created several melody-generation algorithms, including Attention RNN, which introduces an attention mask that allows the RNN to access previous information by relying on more than just the most recent hidden state.

RNNs are less common in modern generation systems since the introduction of Transformer models \citep{og_transformer}, which produced better results compared to RNNs. Transformer models use self-attention to create more consistent, high-quality outputs. To do this, self-attention applies attention weights on the input sequence, rather than attention that was previously common in encoder-decoder systems where weights are applied in the decoding phase onto a separate sequence resulting from encoding. While the original Transformer was created for language tasks, \cite{transformer} adapted the architecture for music, finding that it generated compelling structure even over long sequences. Since then, Transformers have been used widely in deep learning systems for music generation \citep{CompoundWT,gen_via_struct,mingus}.

\cite{transformer} discussed \cite{lattner} suggesting that self-attention serves as a more general form of self-similarity, pointing to the possibility that self-similarity could be used in a similar manner to attention mechanisms to constrain the structure of a piece of music. \cite{lattner} also take advantage of self-similarity matrices (SSMs) by using a loss function that incentivizes self-similarity structure within the music, finding that it improves higher-level structure within the music. 

In many successful methods which produce high-quality structured outputs, the user has little control over the structure of a generated piece. For instance, Transformers’ self attention mechanism is learned by the system, rather than being set by the user; MusicLM \citep{agostinelli2023musiclm} produces music with long-structure over the course of a minute, but the structure of the music is not user-controlled. The methodology used in \cite{lattner} seeks to produce music with structure, but can only replicate the structure of the original piece it samples from. 

The attention mechanism described in this paper bears similarities to some of the ideas behind Transformers \citep{og_transformer,transformer} and builds on the ideas used by \cite{lattner} to incentivize self-similarity structure in music. However, our attention layer allows the user to describe the desired structure of the piece via an input SSM, giving them control over the structure of a piece independent of the piece's associated notes. While we demonstrate its success using an LSTM as its generation method, this system can be combined with any recurrent generative model that returns a probability distribution over the next note in a sequence. When combined with a fast model such as LSTMs, this method allows for long-term structure generation over multiple minutes, longer than MusicLM \citep{agostinelli2023musiclm}.

\section{Methods}
We construct SING, a proof of concept model  with two layers: an \textit{LSTM layer}, which takes in a sequence and outputs its predictions for the next elements of the sequence, and our \textit{attention layer}, which applies pairwise self-similarities from the input SSM as weights to the predictions from the LSTM. We use a large music dataset as examples of existing music to train the model. The process of training updates the weights for the LSTM layer and the attention layer, allowing them to make better predictions. Once the model has been trained, it can output new music given a short sequence of starting vectors and a template SSM.

We choose an LSTM as the generative model in our proof of concept for two reasons. The first is that in practice, LSTMs often struggle to maintain long-term structure, providing a clearer opportunity to demonstrate our attention mechanism's contribution to creating long-term structure. The second is that LSTMs are more efficient than more complex computational models when generating longer pieces, allowing us to generate and evaluate comparatively long pieces.

\subsection{Dataset}
For training, we use the MAESTRO dataset (MIDI and Audio Edited for Synchronous TRacks and Organization dataset) \citep{dataset}, curated by Google Magenta, which comprises MIDI files of classical music from the International Piano e-Competition.  It contains sections of classical music long enough that there would be repeated structure for the network to try to emulate, and is large enough that it was feasible to train a neural network on it with around 160 hours/5.5 million notes of music, with additional data available to validate and test the model. It also presents a significant challenge for an LSTM, as it is \textit{polyphonic} music (i.e. music that has multiple melodic lines playing at once) rather than individual melodies.

\subsubsection{Data Pre-Processing}\label{subsec:batching}
We pre-process the data to be compatible with self-similarity computations by converting the MIDI files into piano roll representation\footnote{We use the \texttt{pretty\_midi}\citep{pretty_midi} library to convert the MIDI files to piano roll.}. After pre-processing, we have 4096 pieces of size 128 pitches by 255 to 700 samples.

 We estimate the tempo of each piece using \texttt{pretty\_midi} and sample at that tempo to generate the piano roll for that piece. Tempo estimation in \texttt{pretty\_midi} is not very accurate on complex music like the MAESTRO dataset, but it does provide a general benchmark of the number of musical events (i.e. note changes) per minute. Over the entire dataset, it estimated the average tempo at 205 events per minute.

We also, for simplicity's sake, convert the piano roll to a binary matrix (which also allows us to use binary loss functions). To flatten the piano roll into a binary matrix, if the velocity is not 0 (i.e., if the note was not silent), we set the value to one; otherwise, it stays at zero. The rationale for this threshhold is that even if a note is quiet, it should still be marked as ``on.''

\subsubsection{Variable-Length Batching}

Batching methods require piece length to be consistent within a batch. As noted previously, traditional batching methods extend or truncate every piece to one standard length. 
Here, we present our novel batching method that uses pieces of varying sizes with minimal exclusion or extension. Our method consists of a padding/truncation scheme combined with uniform-length batching, allowing for long pieces to be broken down into several smaller pieces. 

In our variable batching method, we truncate or extend each piece to one of 16 standard pre-selected lengths. Due to computational limitations, we slice pieces longer than 700 samples (about 3.4 minutes) into multiple shorter segments of equal size. Then, we determine standard piece lengths using an exponential fit, where the first standard length is the $k$-shortest piece (in our case $k=10$, 255 samples) and the sixteenth standard length is the maximum piece length (700 samples). We assign pieces to standard lengths based on closest log distance, such that each piece is truncated or extended by no more than 4\% of its length. This method allows us to batch pieces for efficiency without excessive extensions and/or truncations of length, and we retain structures in our training dataset for a large variation of piece lengths, up to multiple minutes of structure.

\subsubsection{Self-Similarity Matrices (SSMs)}

The target SSM of the original piece is an input for the system. In order to generate the SSM for each piece in the MAESTRO dataset, we calculate the chroma vectors from the MIDI of each sample in the pre-processed piece and then compute pairwise cosine similarity\citep{cos,self-sim}.
For each pre-processed piece of length $n$ samples, we have a SSM of size $n$ by $n$.

\subsection{Network Structure}\label{sec:Model}
The first layer in SING is a generative layer (in this case, an LSTM), which takes in a sequence and outputs a probability distribution of the next element of that sequence. The second is our attention, a linear layer that applies weights calculated from an input SSM to the output of the LSTM layer. 
\subsubsection{LSTM Layer}

The first layer in SING is a LSTM layer that uses both the input sequence and the SSM of the training piece. 
The first step in the generation process is to generate the LSTM's best prediction for the next sample, in the form of a probability distribution for each beat.
The network generates new samples until the it reaches the length of the training piece.
During training, the LSTM's parameters are tuned using gradient descent to optimize the weights the network uses to make its predictions. We use a single-layer LSTM with a hidden size of 128.

\subsubsection{Sparsemax Activation}

In the LSTM, we use the sparsemax function from \cite{sparsemax} as an activation function. This function is similar to the typical softmax function in that it regularizes a distribution so it sums to one, but unlike softmax, sparsemax allows for the possibility of numbers being close to or equal to zero, as long as the entire distribution still sums to one. This preserves the relative weights more accurately. Sparsemax is the Euclidean projection of the input onto the probability simplex. Martin et al.~\citep{sparsemax} define the $K-1$ dimensional probability simplex $\Delta ^{K-1}$   as
 \begin{equation}
     \Delta^{K-1} = \{\bf{p}\in \mathbb{R} |  \sum{p} = 1, \bf{p} \geq \bf{0} \}
 \end{equation}
 and provide the following definition for sparsemax:
 \begin{equation}
     \textrm{sparsemax}({\bf{q}}) = \underset{\bf{p}\in\Delta^{K-1}}{argmin}\|\bf{p-q}\|^2
 \end{equation}

\subsubsection{Attention Layer}

The key idea of SING is to use a SSM to find the attention weights for each timestep. The purpose of attention is to highlight the beats that the network should pay more attention to; the SSM essentially functions as a weighted list of the past beats that are most similar to the beat being generated. Instead of prompting the network to generate its own weights, we use the information from finished pieces as a template for the generated piece to replicate. 

To apply the weights to each layer, we pass the SSM of a training piece to the network during each forward pass. 
Through indexing, the network finds the row of the SSM associated to the current generation step and finds the self-similarity values of the previous elements of the sequence to the currently generated element.

SING then runs these values through a sparsemax layer, and uses this regularized similarity to get a weighted sum of the previous elements. The resulting 128-dimensional vector is concatenated to the output of the LSTM and run through the linear layer using linear regression to combine the two 128-length vectors into one 128-long vector.
 
 More formally, let $y$ denote a sequence of input samples with SSM denoted $S$. At time $t$, we compute the attention vector $a$ for sample $y_t$ as follows. We calculate weights $w$ from $S$ and the attention vector $a$ as:
	\begin{equation}
	    W = \textrm{sparsemax}(S_{t,\{ 1,\ldots,(t-1)\}})
	\end{equation}
	\begin{equation}
	    a = \sum_{i=1}^{t-1}(w_i \times y_i)
	\end{equation}
	where $S_{t,\{1,\ldots,(t-1)\}}$ is the first $(t-1)$ entries of the $t^{th}$ row of $S$. Finally, we apply a linear transformation to $a$ and the LSTM output vector $z$ to get the final distribution vector, $d$:
\begin{equation}
    d_j =\textrm{linear}(a_j, z_j), \text{for }1\leq j \leq 128
\end{equation}
\subsection{Sampling}
To get the next sample of notes in the sequence from this distribution $d$, we use
top-50 sampling \citep{sampling} to sample the model's output and choose up to 3 notes per sample.

Sampling from the output distribution, rather than only picking the highest probability notes, diversifies the selection of the notes \citep{sampling}. We limit the sampling to exclude the lowest and highest 20 pitches in the MIDI scale, as these are rarely used in MAESTRO.

\begin{figure*}[t]
    \centering
    \includegraphics[width=0.99\textwidth]{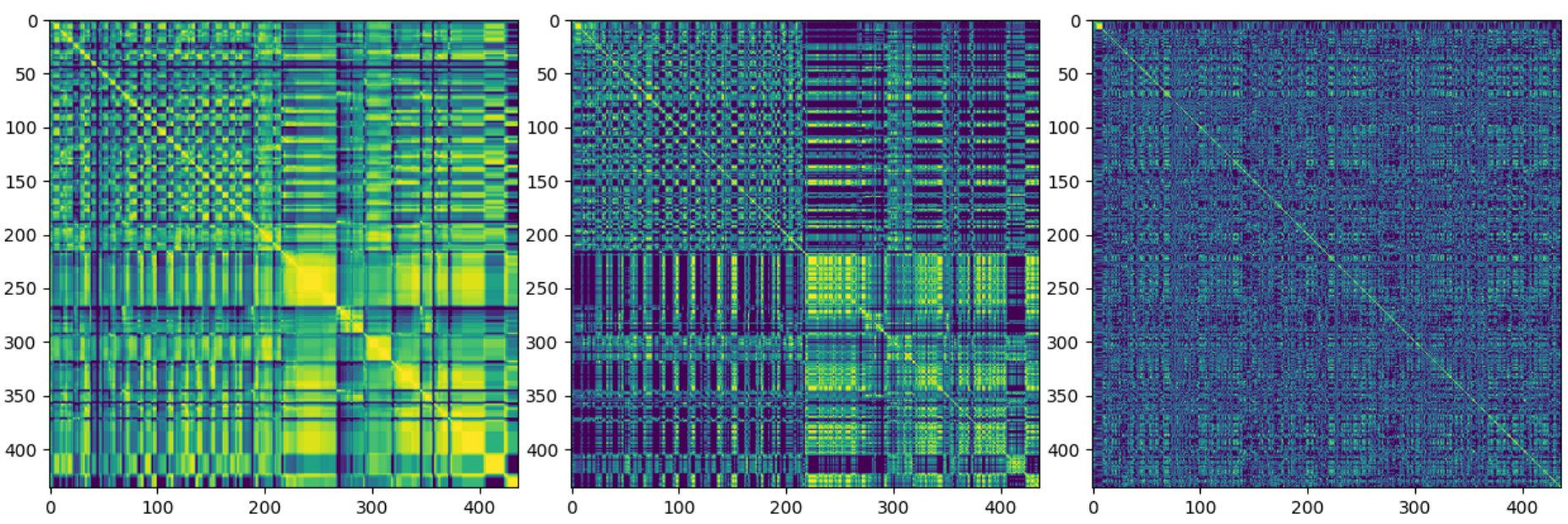}
    \caption{An example of the SSMs from generation. From left to right, the original SSM; the SSM generated by SING; and the SSM generated by an LSTM. The SSM of SING is closer to the original SSM than the comparison model, which demonstrates little structure, if any.}
    \label{fig:generation-results}
\end{figure*}

\subsection{Loss Function}\label{subsec:loss}
It can be challenging to define terms like ``correct,'' ``wrong,'' and ``ground truth'' in the context of music generation, as music does not necessarily have correct or incorrect answers.
In the training step, SING is incentivized to replicate the original piece as closely as possible. In this context, ``ground truth'' is the original piece SING trains on, a ``correct'' note would be activating the same note as in the original piece, and a ``wrong'' note would be incorrectly activating a note when it is not in the original piece. For the purposes of this experiment, replicating a different piece than the input SSM and initial notes would be considered ``incorrect.''

Drawing on \cite{lattner}, our loss function has two components. The first part is the Pytorch implementation of multi-label binary cross-entropy with logits \citep{BCE} (referred to hereafter as BCE loss). Assuming $x_i$ is the $i^{th}$ sample of the generated piece, and that $y_i$ is the $i^{th}$ sample of the training piece, the BCE loss is: 
\begin{equation}
\begin{split}
     l_{BCE}(x_i,y_i) =& \sum_{j=1}^{128}(y_{i,j}\text{log}\sigma(x_{i,j}) + \\
     & ~~  ~~  ~~ (1-y_{i,j})\text{log}(1-\sigma(x_{i,j})))
     \end{split}
\end{equation}

Similarly to the self-similarity constraint proposed by \cite{lattner}, the second part of our loss function computes the mean squared error between the SSMs of the target and generated sequences. This second component encodes the system's loss at the level of the overall structure. If $G$ is the generated self-similarity matrix, $S$ is the target self-similarity matrix, and $n$ is the length of the pieces, the second component of our loss function is: 

\begin{equation}
     l_{MSE}(G,S) =\frac{1}{n^2}\sum_{i=1}^{n}\sum_{j=1}^{n}(G_{i,j}-S_{i,j})^2
\end{equation}

Our final loss function computed for each generated piece of music is therefore:
\begin{equation}
     l(x,G,y,S) = \sum_{i=1}^{n}l_{BCE}(x_i,y_i) + l_{MSE}(G,S)
 \end{equation}

\begin{table*}[t]
    \centering{
    \begin{tabular}{c | lc}
    \hline
    \hline
    \textbf{Statement}&\textbf{Comparison }&\textbf{SING Performance}\\
    \hline
    
    & Random &\textbf{+7.03\%}  \\ 
    {\parbox{7cm}{``This piece was interesting.''}} & LSTM &\textbf{+5.52\%} \\
    & Original  &\textbf{-10.56\%}  \\
    \hline
    
    & Random & \textbf{+7.89\%}\\ 
    {\parbox{7cm}{``This composer seems proficient at composing Western classical music.''}} & LSTM  & +3.80\%  \\
    & Original &\textbf{-22.22\%} \\  
    \hline
    
    & Random & -1.89\% \\ 
    {\parbox{7cm}{``This piece sounds like an expressive human performance.''}} & LSTM & +4.06\%\\
    & Original &\textbf{-18.81\%}  \\  
    \hline
   
    & Random  & \textbf{+16.94\%}  \\ 
    {\parbox{7cm}{``I like this piece.''}}& LSTM & \textbf{+9.55\%}  \\
    & Original  &\textbf{-27.70\%} \\  
    \hline
    \hline
    \end{tabular}
    \caption{The results from the human evaluation study comparing 30-second pieces from SING to 30-second pieces from other methods (pieces generated by a base LSTM, pieces generated from uniform random noise, and the original pieces by human composers). For each pairwise comparison, we show the percentage change in scores with SING. Statistically significant scores (with $\alpha=0.05$ in a single-tailed paired t-test) are bolded. For all comparisons we have a sample size of at least 146 ratings.}
    }
    \label{tab:human}
\end{table*}
\subsection{Training Process}
SING is trained on batches using our variable batching system (Section~\ref{subsec:batching}). Each piece in the batch is used as an input to generate a new piece which is then compared back to the original piece via our loss function.
For each training piece, the first ten samples are fed as input into the model along with the SSM. The model returns a probability distribution for the next time step, which is sampled and those samples are appended to the generated sequence. 

Our network chooses randomly whether to append the generated samples as the input for the next forward pass, or to use the original samples from the input piece at that time step. \cite{sched_sample} use a scheduled rate of decay in probability; we consistently use $p = 0.8$ as the probability of using the generated samples. 
If in contrast, the model were trained autoregressively on its own outputs, the network will often generate inaccurate probability distributions early in training and thus predictions will degenerate rapidly; if it is trained only on the correct data, the model will suffer from exposure bias and may struggle at generation time when it has to train on its own outputs. 
Choosing randomly whether to use the generated samples or the original piece's samples keeps the network on track while reducing the train/test mismatch.

The forward pass repeats until the generated sequence reaches the original piece's length.
This output sequence is then run through the loss function described above and the network adjusts weights using gradient descent.
We use the Adam optimizer \citep{adam} and a learning rate of 0.001.
Once trained, SING can be used to generate new pieces of music as piano roll. That piano roll can then be converted back into listenable audio.

\subsection{Methods of Evaluation}\label{subsec:eval_methods}
We evaluate SING in two ways, addressing two different definitions of ``better performance'' as a music generation system. First, we conduct a human evaluation study to examine the comparative musicality of the output against two comparison models. In our second evaluation, we leverage standardized mean squared error (MSE) to quantify structural similarity of the generated output to the template SSM. In the first evaluation we focus on the musicality of the generated pieces, while in the second, we seek to computationally quantify the amount of structure in the newly generated piece. 

To qualify the musicality of SING output, we conducted a human evaluation study rating short clips generated by SING. The rationale behind the survey is to determine whether SING's improvements to long-term structure also translate to improvements in musicality of the generated output. We distributed an IRB-approved study to university students and received 61 responses.\footnote{We had 79 students start the survey but only 61 people answered beyond the demographic information.} Participants listened to 8 pairs of 30-second samples, where one sample was generated by SING, and the other sample was one of three controls:  the comparison model, random noise from a uniform distribution, or the original composition from which SING replicated structure, drawing on survey methods from \cite{gen_via_struct} and \cite{ chi2020generating}. Participants rated samples on a Likert scale (1=Strongly Disagree; 5=Strongly Agree) on four dimensions of musicality: interest, proficiency, expressiveness, and likeability, similar to categories used by \cite{CompoundWT}. We use a one-sided paired $t-$test with $\alpha = 0.05$ to test whether SING scored significantly greater, or significantly worse, than each control. For each pair, participants also rate whether samples seemed similar. For this test, we used a $t-$test for scores related by participant. 

To quantify the structural similarity of the output pieces, we report the MSE between the standardized template SSM and standardized generated SSMs. The rationale behind standardizing the SSM is that it emphasizes the structure of the SSM, rather than the literal values; in practice, zero-mean/standardized metrics are commonly used in template-matching applications \citep{NAKHMANI2013315}. We report the average standardized MSE for SING, a basic LSTM, and random noise. For each method, we generate three SSMs for all 438 processed pieces in the test set; thus, we report the average standardized MSE over 1314 SSMs.

\section{Results and Discussion}

Two music generation systems were successfully trained on the MAESTRO data. The first was SING as described in Section~\ref{sec:Model}, using our proposed attention mechanism; the second was an ablated comparison with the same number of layers and hidden size as SING, but with the attention mechanism removed. We choose this comparison to pinpoint the efficacy of our proposed attention mechanism. We also report statistics for music created by uniform random sampling.

\subsection{Training Models using Variable-Length Batching}

The networks were trained on MAESTRO data. In the MAESTRO dataset of classical music, piece lengths vary from 103 to 9156 samples, according to our sampling method. Piece lengths are skewed right, with multiple long pieces outside three standard deviations above the mean length, such that any single standard length ill-captures the lengths of pieces in this dataset. After pre-processing our data, the training set contains 3262 pieces of 255 to 700 samples each, where an individual sample contains binary data on the on/off state of 128 pitches. 

Each network was trained using batching with up to 100 pieces per batch. 
Since piece length must be consistent within a batch, pieces were grouped by length before batch assignment, resulting in 38 batches of varying size. The length restriction prevents completely random batching, but batch assignment and order was otherwise random.

\begin{figure*}[t]
    \includegraphics[width=0.99\textwidth]{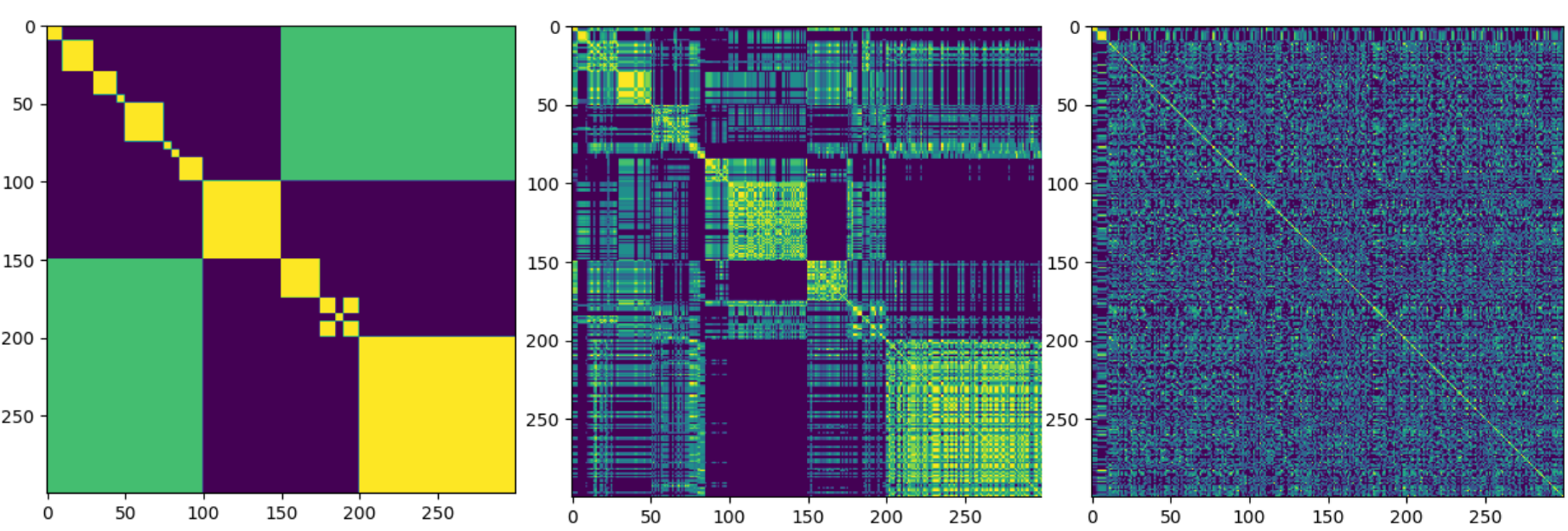}%
    \caption{From left to right, a synthetic self-similarity matrix, the piece as generated by SING, and the piece as generated by the comparison model. SING generates a piece that resembles the synthetic SSM, while the comparison model cannot.}
    \label{fig:noise-results}
\end{figure*}

\subsection{Selecting Best Models}

For both SING and the comparison LSTM, we select the best model (associated to a particular number of epochs), by validating each epoch-labelled model on pieces from the MAESTRO validation set. For both SING and the comparison model, we picked the model with the best performance on the validation set as the model to evaluate. SING achieved its best performance on the validation dataset at the 24th epoch, while the base LSTM performed best at the 3rd epoch.\footnote{We report training efficiency on a single 32 GB GPU: The comparison model took 13 hours to train for 30 epoch, and training SING under the same conditions took 22 hours.}

\subsection{Evaluation Results}
Our first evaluation concerns the musicality of SING-generated pieces when compared to a simple LSTM and two additional controls: random noise and the original piece from which SING copied structure. The results of the human evaluation study can be found in Table 1. When asked if the two pieces of music the participants were comparing were similar, participants generally agreed samples were similar for all comparisons; we therefore do not report those results. According to our statistical tests, SING scored significantly higher than the comparison model on interest and likeability but showed no statistically significant difference in perceived proficiency at composing Western classical music. SING generally outperformed random noise, but performed worse than the original piece. Therefore, SING’s attention mechanism improves on musicality over the comparison model, but falls short of improving the perceived skill of the composer. SING’s perceived expressiveness was not significantly better nor worse than random noise. The results of our study suggest that SING’s attention mechanism improves interest and likeability over a basic LSTM, but does not necessarily demonstrate significant improvements in perceived musicality for short clips of SING-generated pieces.

Our second evaluation compares the SSMs of the template piece and the generated piece. The standardized MSE is significantly lower for SING compared to the LSTM and random noise (Table \ref{tab:results}), indicating that our method largely succeeds in matching the structure of the template piece.

\begin{table}[t]
    \centering{
    \begin{tabular}{c|c}
    \hline
    \hline
    \textbf{Generator}&\textbf{MSE} \\
    \hline
        random& 1.96\\
        LSTM& 1.96\\
        \texttt{SING}& 1.57\\
    \hline
    \hline
    \end{tabular}
    }
    \caption{Mean squared error of the standardized SSMs (comparing structural similarity to template piece). Lower MSE corresponds to better similarity. }
    \label{tab:results}
\end{table}

\subsection{Generation Results on Music}

Figure~\ref{fig:generation-results} shows an example of the SSMs for the generated outputs from a basic LSTM and from SING compared to the original SSM for the input piece. It is evident in these cases that the proposed SSM-based attention component does incentivize the network to generate a similar structure to the input SSM, while the comparison model (without attention) struggles to emulate it at all. The attention mechanism replicates the larger elements of structure much better than smaller-scale elements of structure.

\subsection{Generation Results on Artificial SSM}

To further test the capabilities of the network in replicating SSMs, we here provide an example of the network replicating an artificially created self-similarity matrix. We create the synthetic SSM seen in Figure \ref{fig:noise-results}. SING generates a piece that resembles the artificial SSM, unlike the comparison model. This demonstrates SING's ability to replicate SSMs even when they are manmade, and not similar to SSMs derived from existing pieces.

\section{Future Work and Conclusions}

This paper introduces an attention mechanism that uses self-similarity matrices (SSMs) as an input to function as attention to create long-term structure over the course of the piece. We provide a demonstration of its efficacy using Similarity Incentivized Neural Generator (SING), a deep learning music generation system. Overall, there is evidence that the proposed attention mechanism is effective at replicating long-term structure in music over the course of the piece, even on long timescales of up to 700 beats (approximately 3 minutes). Applying an SSM as attention to a generation system is an effective way to incentivize the generation system to build music with a given structure.

The variable-length batches used to train SING present a data processing solution for musical datasets with large variation in piece length, minimizing the need for piece extension or truncation for batching. With more computational power, this batching method could enable training on music even longer than 700 beats, for the replication of structure on extremely long scales.

As previously mentioned, Transformers can often achieve state-of-the-art results in generation; it would be interesting to create a system that uses Transformers as a generative model for our proposed attention layer. This may give users control over the structure of a piece while gaining the advantages of the typically higher quality generations of the Transformer, potentially addressing the limitations seen in the user evaluation.

The model demonstrating our attention mechanism, SING, is limited by its sampling method, which requires at least one note to be activated at any time. Improving the sampling to take into account the possibility of silence could add to the network's capability to generate similar structures, and could improve the listenable audio. Similarly, the maximum number of notes allowed on in this model was three; future work might take into account the possibility of more notes being simultaneously played.

Another limitation is that SING requires an existing SSM as input, which both impacts the generation step and the loss function. SING appears to already be relatively consistent when generating large-scale structure, but may be being penalized for deviations from smaller structure elements; the thresholded and ``blurred'' SSM proposed by \cite{Grosche} might be more forgiving, allowing systems which use our proposed attention mechanism to learn larger elements of structure more reliably.

Our data processing relies on imperfect automated tools to determine tempo, and we further alter the data by binarizing it. Our system is limited by the reliance on these alterations; training on beat-annotated or non-binarized data could improve the quality of the output music substantially.

Our proposed attention mechanism demonstrates promising ability to generate music with a user-suggested structure on longer time-scales. Additionally our method generates music without the same training needs as a Transformer. We recognize that improvements to the quality of the proposed system's outputs could be made in a variety of ways, including different training data, more complex generation layers, or improved sampling methods. Nonetheless,  this paper's proposed ability to control structure may be useful in deployment of future music generation systems, potentially allowing users of systems with similar mechanisms to freely alter and compose music to their requirements without requiring extensive knowledge of composition techniques.

%%%%%%%%%%%%%%%%%%%%%%%%%%%%%%%%%%%%%%%%%%%%%%%%%%%%%%%%%%%%%%%%%%%%%%%%%%%%%%%%
% Please do not touch.
% Print Endnotes
\IfFileExists{\jobname.ent}{
   \theendnotes
}{
   %no endnotes
}
%%%%%%%%%%%%%%%%%%%%%%%%%%%%%%%%%%%%%%%%%%%%%%%%%%%%%%%%%%%%%%%%%%%%%%%%%%%%%%%%

%\section*{Acknowledgements}

%Any acknowledgements must be headed and in a separate paragraph,
%placed after the main text but before the reference list.

%%%%%%%%%%%%%%%%%%%%%%%%%%%%%%%%%%%%%%%%%%%%%%%%%%%%%%%%%%%%%%%%%%%%%%%%%%%%%%%%
% Bibliography
%%%%%%%%%%%%%%%%%%%%%%%%%%%%%%%%%%%%%%%%%%%%%%%%%%%%%%%%%%%%%%%%%%%%%%%%%%%%%%%%

% For bibtex users:
\bibliography{ISMIRtemplate}

% For non bibtex users:
%\begin{thebibliography}{citations}
%
%\bibitem {Author:00}
%E. Author.
%``The Title of the Conference Paper,''
%{\it Proceedings of the International Symposium
%on Music Information Retrieval}, pp.~000--111, 2000.
%
%\bibitem{Someone:10}
%A. Someone, B. Someone, and C. Someone.
%``The Title of the Journal Paper,''
%{\it Journal of New Music Research},
%Vol.~A, No.~B, pp.~111--222, 2010.
%
%\bibitem{Someone:04} X. Someone and Y. Someone. {\it Title of the Book},
%    Editorial Acme, Porto, 2012.
%
%\end{thebibliography}

\end{document}